\documentclass[aps,prl,twocolumn,floatfix,superscriptaddress,amsmath,amssymb]{revtex4-2}

\usepackage{graphicx}
\usepackage{bm}
\usepackage[pdftex]{color}
\usepackage{xcolor}
\usepackage{booktabs}
\usepackage{hyperref}
\usepackage[mathlines]{lineno}
\usepackage{braket}
\usepackage{bbm}
\usepackage{verbatim}
\usepackage{ulem}
\usepackage{standalone}
\usepackage{comment}
\usepackage[height=8.85in,width=6.45in]{geometry}
\usepackage{array}
\usepackage{slashed}
\usepackage{braket}
\usepackage{tikz}
\usepackage{tikz-cd}
\usepackage{times}
\usepackage{courier}
\usepackage{bm}
\usepackage{xcolor}
\usepackage{mdframed}
\usepackage{datetime}
\usepackage{dsfont}
\usepackage{multirow}
\usepackage{cancel}
\usepackage{caption}
\usepackage{subcaption}
\usepackage{graphicx}
\usepackage[compat=1.1.0]{tikz-feynman}
\usepackage{pifont}
\usepackage{array} 
\usepackage{standalone}

\newcommand{\parR}[1]{\textbf{\textit{#1}}---}

\newcommand{\Z}{\mathbb{Z}}

\renewcommand\[{\begin{equation}}
\renewcommand\]{\end{equation}}

\def\ie{\begin{equation}\begin{aligned}}
\def\fe{\end{aligned}\end{equation}}

\begin{document}

\title{Symmetry-Enforced Topological Structures in Quantum Phase Diagrams}

\author{Linhao Li}
\affiliation{Department of Physics, the Pennsylvania State University, University Park, Pennsylvania 16802, USA}

\author{Yuan Yao}
\email{smartyao@sjtu.edu.cn}
\thanks{Corresponding author.}
\affiliation{School of Physics and Astronomy, Shanghai Jiao Tong University, Shanghai 200240, China}

\date{\today}

\begin{abstract}
We study the topological structure of the quantum phase diagram of gapped systems by identifying the noncontractibility of loops,
so-called $S^1$-families,
within the gapped phase diagram in which many-body Hamiltonians can have nontrivial ground-state degeneracy.
We manifest the role of symmetries in such $S^1$-family classifications by the exotic symmetry interplay:
(i) nontrivial mixed anomalies, (ii) semi-direct product relation between the spontaneously broken and the unbroken symmetries,
and (iii) symmetry with noninvertible operators.
We find that such structures lead to the novel $S^1$-family classifications
inaccessible by earlier classifications based on ``independent'' symmetries.
Furthermore, we construct lattice realizations of these $S^1$-families and explicitly demonstrate their novel algebraic structures.
\end{abstract}
\maketitle

\begin{figure*}[t]
    \centering
    \includegraphics[width=1\linewidth]{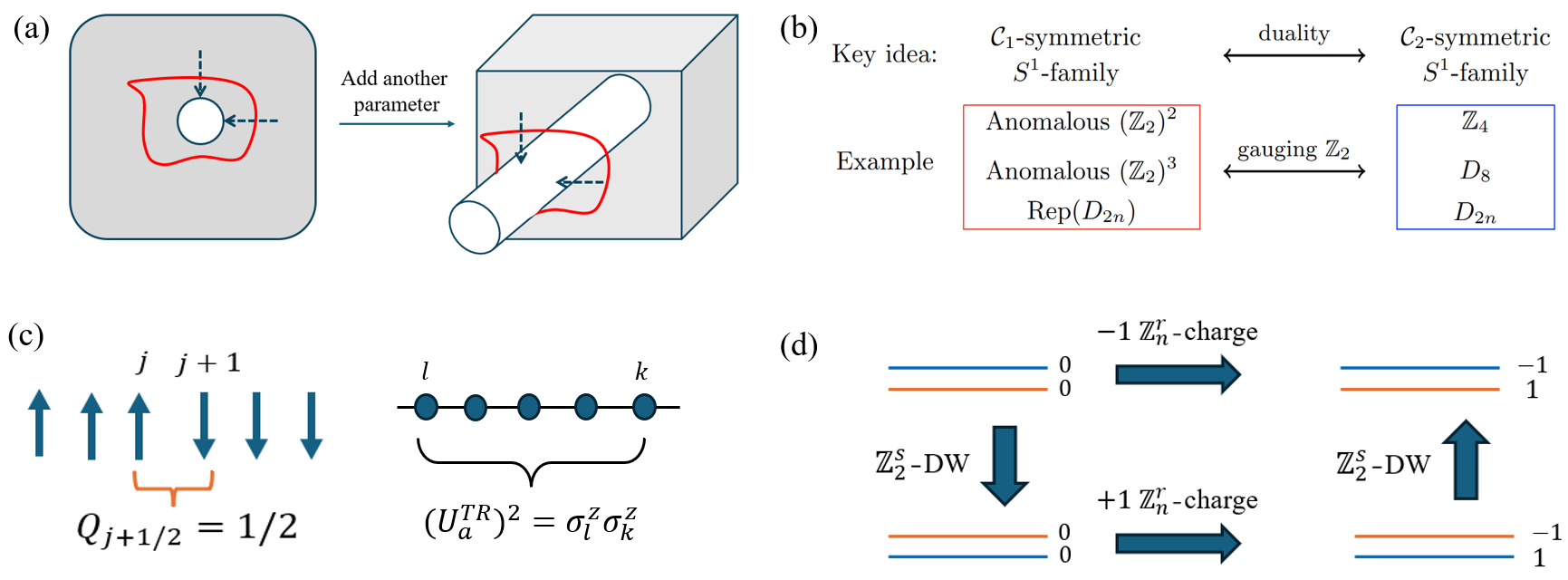}
    \caption{(a) Noncontractible loops are stable, e.g., adding another parameter to the phase diagram cannot trivialize the non-contractibility {due to the inevitable gapless phase (white area) bounded by the loop. In the current case, we need to finetune only two parameters to reach the gapless phase numerically or experimentally.} 
    (b) The duality method to study $S^1$-family  with examples shown in the boxes. (c) For the $\mathbb{Z}_2^x\times\mathbb{Z}_2^a$ symmetry, a domain wall of either $\mathbb{Z}_2$ symmetry carries a half charge of the other. Here, $U_a^{\mathrm{TR}}$ denotes the truncation of $U_a$ to a finite interval $l\leq j\leq k$, which creates a $\mathbb{Z}_2^a$ domain wall at each boundary. 
The second panel therefore means that two $\mathbb{Z}_2^a$ domain walls carry one $\mathbb{Z}_2^x$ charge. (d) A combined evolution of $\Z^s_2$ domain-wall pumping and $\Z^r_n$ charge pumping. Here the blue and orange lines label the ground state of Eq.~\eqref{eq: Z2SSB main text} with $\tau^z=\pm 1$ respectively and numbers on the right label the pumped $\Z^r_n$ charge associated with each ground state. }
    \label{fig:type-II anomaly}
\end{figure*}

\parR{Introduction.}
Understanding quantum phases and phase transitions is a central challenge in many-body physics.
The gapped-phase diagram is partitioned into disconnected parts,
among which phase transitions are inevitable.
Two gapped systems belong to the same phase if they stay in the same partition. 
The identification of all these phases is called the classification of gapped phases including spontaneous-symmetry-breaking (SSB) \cite{Landau:1937aa},
topological-ordered \cite{Wen1995,kitaev2003fault,PhysRevLett.96.110405,PhysRevLett.96.110404} and symmetry-protected topological (SPT) phases \cite{PhysRevB.80.155131,PhysRevB.81.064439,PhysRevB.85.075125,PhysRevB.83.035107,PhysRevB.84.235128,PhysRevB.84.235141,PhysRevB.87.155114}.
Furthermore,
recent progress has been made on
more intrinsic structures,
e.g., noncontractible loops called nontrivial $S^1$-family of gapped Hamiltonians \cite{Kitaev2011SRE,Kitaev2013SRE,PhysRevB.102.245113,PhysRevB.101.235130,SciPostPhys.8.1.001,PhysRevB.106.165115,PhysRevB.110.035114,PhysRevB.106.125108,bachmann2023classificationgchargethoulesspumps,PhysRevLett.129.017204,PhysRevB.110.094410,PhysRevB.108.125147,inamura20241+,li2025classification,inamura2026generalized,ohyama2026parameterized,manjunath2026search,ohyama2026parameterized};
continuously shrinking a certain loop to a single point \textit{within the gapped phase diagram} is impossible~\footnote{In the entire phase diagram,
such a contraction is possible in the price of inevitably touching a gapless point,
extending the concept of inevitable phase transitions before.} even after arbitrarily more experimental parameter axes are introduced as in FIG.~\ref{fig:type-II anomaly}~(a). {It provides a topological mechanism that one needs to fine-tune only a finite number of parameters to achieve certain gapless phases, even though \textit{the} quantum phase diagram has infinitely many axes.}

Symmetry-preserving and SSB $S^1$-families have been established,
separately \cite{Kitaev2011SRE,Kitaev2013SRE,PhysRevB.102.245113,bachmann2023classificationgchargethoulesspumps,PhysRevLett.129.017204,PhysRevB.110.094410,inamura2026generalized}.
The situation of $G$-SSB $S^1$-family with an unbroken $K$ symmetry,
i.e., partially SSB,
could be {naively} analyzed by hybridizing these two approaches,
which is clearly true if the unbroken and broken symmetries are independent,
e.g., $G$ and $K$ are onsite and commute on each site. 
However,
it is quite often that,
the SSB is a compulsory consequence due to some \textit{nontrivial interplay} between $G$ and $K$.
The famous Lieb-Schultz-Mattis theorem \cite{Lieb:1961aa} is one notable example,
where spin-rotation symmetry and translation symmetry have a mixed anomaly \cite{Affleck:1986aa, OYA1997,Oshikawa:2000aa, Hastings:2004ab,Fuji-SymmetryProtection-PRB2016,Watanabe:2015aa,Ogata:2018aa,Ogata:2020aa,Yao:2021aa,Affleck:1986aa,Yao:2019aa,PhysRevB.110.045118,PhysRevB.106.224420,PhysRevLett.133.136705,seiberg2025lsm,aksoy2024lieb}.
In addition,
$G$ and $K$ may have nontrivial action to each other,
which is common in systems with both charge-conjugation and charge-conservation  symmetry.
Furthermore,
the interesting symmetry nowadays has been generalized to include noninvertible operators ~\cite{PhysRevLett.93.070601,Frohlich:2006ch,
aasen2016topological,aasen2020topological, inamura2022lattice,thorngren2024fusion,thorngren2024fusion,Chen:2023qst, Lootens_2023, Lootens_2024, Eck_Fendley_1,PhysRevB.107.125158, SciPostPhys.18.5.153,Li_2023, Cao:2023doz,Seiberg:2023cdc, Sinha:2023hum, Seiberg:2024gek,g1l7-bvtf, Seifnashri:2023dpa, tavares2025integrable, Cao:2024qjj,Lu:2024ytl,Pace:2024tgk, SciPostPhys.18.1.028, SciPostPhys.18.4.121, Okada:2024qmk, PhysRevB.111.115112,SciPostPhysCore.8.4.070,249m-m8wq,SciPostPhys.18.1.032,PhysRevLett.133.161601,SciPostPhys.18.1.008,li2024non,PhysRevB.107.205137,wen2026non},
and
$S^1$-families on such symmetric phase diagrams are largely unknown.
Therefore,
how to identify and classify the nontrivial $S^1$-family in the presence of the above nontrivial symmetry properties remains an open but essential question.

Quite often,
exotic and complicated symmetry structures can be unravelled from another viewpoint with a more transparent physical picture,
by so-called \textit{duality} methods,
a general technique to study phase classifications \cite{Kramers-Wannier,RevModPhys.51.659,PhysRevB.45.304,Kennedy:1992tke,oshikawa1992hidden,PhysRevB.90.035451,PhysRevLett.112.141602,Lootens_2023,Li_2023,SciPostPhys.18.5.156,SciPostPhys.19.4.113}.
In this \textit{Letter},
we apply the duality approach to investigate $S^1$-families with the above exotic symmetry interplay.
We show that the mixed-anomaly or semidirect-product structure between unbroken $K$ and broken $G$ symmetries intertwines the noncontractible loops associated with $K$-charge and $G$-domain-wall (c.f., discussions later), leading to novel $S^1$-family classifications. 
We further classify the $S^1$-families of $\operatorname{Rep}(D_{2n})$ noninvertible SPT (NISPT) phases \cite{Seifnashri:2024dsd,4zfz-x9xh,inamura20241+,li2025classification,inamura2026generalized,ohyama2026parameterized},
where the $S^1$-family classification has an intrinsic base-point dependence,
in contrast to conventional SPT with group symmetry \cite{Kitaev2011SRE,Kitaev2013SRE,PhysRevB.110.094410}.
As concrete illustrations, 
we explicitly construct symmetric unitary evolutions generating the proposed noncontractible loops in lattice models and directly reveal their associated algebraic structures.

\parR{Preparations: KW duality and $S^1$-family of $\Z_2$-SSB phases.}  A duality between symmetries $\mathcal C_1$ and $\mathcal C_2$ establishes a one-to-one correspondence between phase-diagram or so-called parameter space structures and hence between the noncontractible loops on two sides, as illustrated in FIG.~\ref{fig:type-II anomaly}(b) \footnote{For $D_8=\mathbb Z_4^r\rtimes\mathbb Z_2^s$ symmetry, gauging $\mathbb Z_2^s$ gives rise to $\operatorname{Rep}(D_8)$ symmetry, whereas gauging $\mathbb Z_2^{r^2}=\{1,r^2\}$ yields an anomalous $(\mathbb Z_2)^3$ symmetry.}.
We first demonstrate this framework with the simplest example: the Kramers-Wannier (KW) duality, which maps the 1D $\mathbb Z_2$-SSB phase to the $\mathbb Z_2$-symmetric phase by exchanging domain walls (DWs) and charge excitations. Consequently, it establishes a correspondence between the $S^1$ families of these two phases. As shown in Refs.~\cite{PhysRevLett.129.017204,PhysRevB.110.094410}, the $S^1$ family of the $\mathbb Z_2$-symmetric phase is classified by $\mathbb{Z}_2$, which implies the same $\mathbb Z_2$ classification for the $\mathbb Z_2$-SSB phase.

To elucidate the physical meaning of the corresponding noncontractible loops, we consider a spin-$1/2$ chain of length $L$ under periodic boundary conditions (PBC) with $\mathbb Z_2^x$ symmetry generated by $\prod_{j=1}^L\sigma^x_j$. The KW duality transformation acts as
\begin{equation}
    \mathcal{N}:\sigma^z_{j-1}\sigma^z_{j}\to\tilde{\sigma}^x_j, \sigma^x_{j}\to \tilde{\sigma}^z_{j}\tilde{\sigma}^z_{j+1},
\end{equation}
where the (tilded) dual system preserves a dual $\Z^x_2$ symmetry generated by $\prod^L_{j=1} \tilde{\sigma}^x_j$.

{Let us consider a convenient construction of the noncontractible loops in the spin-$1/2$ chain from a  symmetric unitary evolution operator $\mathcal V(\theta)$ with $\theta\in[0,2\pi]$ and $\mathcal{V}(0)=\mathbb{I}$.
It generates noncontractible loops of Hamiltonian through $
H(\theta)=\mathcal V(\theta)H(0)\mathcal V^\dagger(\theta)$, where $H(0)$ is any Hamiltonian in the certain phase within interest, and the loop periodicity condition $H(0)=H(2\pi)$ requires that $\mathcal V(2\pi)$ commutes with $H(0)$.}

For the $\mathbb{Z}^x_2$-symmetric phase, the corresponding evolution unitary is \cite{li2025classification}
\begin{equation}
    \mathcal{V}_{\text{charge}}(\theta)=\exp\left(\frac{ i\theta}{4} \sum^L_{j=1} (1-\sigma^z_{j-1}\sigma^z_j)\right),
\end{equation}
satisfying $\mathcal{V}_{\text{charge}}(2\pi)=\mathbb{I}$ under PBC. 
The nontriviality of this $S^1$-family can be manifested by the associated charge pumping upon truncating (TR) the evolution to a finite interval $l\leq j\leq k$, yielding
\begin{equation}
\mathcal{V}^{\text{TR}}_{\text{charge}}(2\pi)
=\sigma^z_l\sigma^z_k.
\end{equation}
Thus, the truncated evolution creates a $\mathbb{Z}_2^x$-charge at each boundary since $\mathcal{V}^{\text{TR}}_{\text{charge}}(2\pi)\sigma^x_{l,k}\mathcal{V}^{\text{TR}\dagger}_{\text{charge}}(2\pi)=-\sigma^x_{l,k}$, explicitly demonstrating the bulk pumping of $\mathbb Z_2^x$ charge during the evolution.
That is why we use the subscript ``$_\text{charge}$'' above.
This $\mathbb{Z}^x_2$-charge pumping serves as a topological invariant characterizing the $\mathbb Z_2$ classification;
nontrivial charge pumping corresponds to a nontrivial $S^1$-family while trivial pumping corresponds to a trivial one.

Under the KW duality, the dual evolution unitary is
\begin{equation}\label{kw_0}
    \tilde{\mathcal{V}}_{\text{DW}}(\theta)=\exp\left(\frac{i\theta }{4} \sum^L_{j=1} (1-\tilde{\sigma}^x_{j})\right).
\end{equation}
At $\theta=2\pi$, this operator reduces to the dual $\mathbb Z_2^x$-symmetry operator and therefore exchanges the two degenerate SSB ground states.
The truncated operator $\tilde{\mathcal{V}}_\text{DW}^\text{TR}(2\pi)\propto\prod_{j=l}^{k}\sigma^x_j$ creates a $\mathbb Z^x_2$ domain-wall at the boundary of the interval since $\tilde{\mathcal{V}}_\text{DW}^\text{TR}(2\pi)(\sigma^z_{l-1}\sigma^z_l)\tilde{\mathcal{V}}_\text{DW}^{\text{TR}\dagger}(2\pi)=-\sigma^z_{l-1}\sigma^z_l$ and $\tilde{\mathcal{V}}_\text{DW}^\text{TR}(2\pi)(\sigma^z_{k}\sigma^z_{k+1})\tilde{\mathcal{V}}_\text{DW}^{\text{TR}\dagger}(2\pi)=-\sigma^z_{k}\sigma^z_{k+1}$. Thus, the noncontractible loop of the $\mathbb Z^x_2$-SSB phase pumps a $\mathbb Z_2^x$ domain wall, which is dual to the charge pumping realized by the noncontractible loop of the $\mathbb Z_2^x$-symmetric phase before and serves as a  $\mathbb Z_2$ topological invariant.

This discussion generalizes straightforwardly to the 1D $\mathbb Z_n$-symmetric and $\mathbb Z_n$-SSB phases. Their $S^1$ families are both classified by $\mathbb Z_n$, corresponding respectively to charge pumping and domain-wall pumping, and one-to-one correspond to each other under the $\mathbb Z_n$ generalization of the KW duality.

\begin{table}[!tbp]
		\centering
		\begin{tabular}{c@{\hskip 0.2in} c@{\hskip 0.2in} c c c c }
			\hline\hline\\[-1em]
			Total Symmetry                       & unbroken symmetry & $S^1$-family \\
			\hline\\[-1em]	
		       $(\Z_2)^2$                      & $\Z_2$&$\Z_4$  \\
			\\[-1em]
			 &  $\Z_1$ &  $(\Z_2)^2$\\
			\\[-1em]
			\hline\\[-1em]

			$(\Z_2)^3$ & $(\Z_2)^2$ &$(\Z_2)^2$\\
			\\[-1em]
			   & $\Z_2$ & $D_8$\\
			\\[-0.7em]
			~  &    $\Z_1$& $(\Z_2)^3$\\
						\\[-1em]
			\hline\\[-1em]
$D_{2n}=\Z^r_n\rtimes\Z^s_2$ & $\Z^r_n$ &$D_{2n}$\\
			\\$D_{8n}=\Z^r_{4n}\rtimes\Z^s_2$
			   & $\Z^{r^2}_{2n}\rtimes \Z^{rs}_{2}$ & $D_8$\\
			\\$D_{8n+4}=\Z^r_{4n+2}\rtimes\Z^s_2$
			~  &     $\Z^{r^2}_{2n+1}\rtimes \Z^{rs}_{2}$& $(\Z_2)^2$\\\\[-1em]
\hline\hline\\[-1em]
			Total Symmetry                       & NISPT & $S^1$-family \\
			\hline\\[-1em]	
			 Rep($D_{2n}$)&Trivial&$D_{2n}$\\
			\\[-1em]
			
				Rep($D_{8n}$)&Nontrivial&$D_8$\\
			\\[-1em]
			Rep($D_{8n+4}$)&Nontrivial&$(\Z_2)^2$\\
			\\[-1em]
			\hline\hline
		\end{tabular}
\caption{The $S^1$-families of different 1D gapped phases. The first part includes the partially SSB phases of $(\mathbb{Z}_2)^2$ and $(\mathbb{Z}_2)^3$ symmetry, both of which carry a mixed anomaly, as well as the partially SSB phases of $D_{2n}$ symmetry. 
Here, $\mathbb{Z}_2$ and $(\mathbb{Z}_2)^2$ in the unbroken symmetry refer to one or two $\mathbb{Z}_2$ factors of the total symmetry and the trivial unbroken group $\Z_1$ denotes the fully SSB phase. The second part includes the $\mathrm{Rep}(D_{2n})$ NISPT phases,
where “Trivial” and “Nontrivial” refer to the NISPT phase where $S^1$ loops stay.}\label{table1}
	\end{table}

\parR{$S^1$-family of gapped phases with anomalous $(\Z_2)^2$ symmetry}
We now consider cases with a mixed anomaly between unbroken and broken symmetries. Such mixed anomalies induce a nontrivial interplay between domain walls of the broken symmetry and charges of the unbroken symmetry, leading to novel algebraic structures of the corresponding $S^1$ families. As a representative example, we focus on a system with $\mathbb Z_2^a\times\mathbb Z_2^x$ symmetry carrying a mixed anomaly, where the $\mathbb Z_2^a$ symmetry is generated by
 \begin{equation}\label{eq: Z2 anomalous gen}
    U_a =\exp\left(\pi i\sum^L_{j=1}Q_{j+\frac{1}{2}}\right), 
\end{equation}
with $ Q_{j+\frac{1}{2}}= (1-\sigma^z_{j}\sigma^z_{j+1})/4$ counting the $\Z^x_2$ domain wall between two sites. Such a mixed anomaly implies that a domain wall of one $\Z_2$ symmetry carries a half charge of the other \cite{PhysRevLett.114.031601,PhysRevB.91.035134,SciPostPhys.15.2.051,SciPostPhys.14.2.012,PRXQuantum.6.010347}, as shown in FIG.~\ref{fig:type-II anomaly}(c). Moreover, as a consequence of this anomaly, any gapped phase with $\mathbb Z_2^a\times\mathbb Z_2^x$ symmetry must spontaneously break one or both of the two $\mathbb Z_2$ symmetries \cite{Seifnashri:2023dpa}.

We first consider the $\Z^a_2$-SSB phase where only $\mathbb{Z}_2^a$ is broken, i.e., unbroken subgroup $K=\Z^x_2$. Under KW duality upon $\mathbb{Z}_2^x$, the $\mathbb{Z}_2^a\times\mathbb{Z}_2^x$ symmetry is mapped to a $\mathbb{Z}^r_4$ symmetry generated by
\begin{equation}
    \tilde{U}_{r}=\exp\left(\frac{\pi i}{4} \sum^L_{j=1} (1-\tilde{\sigma}^x_{j})\right).
\end{equation} 
The physical intuition of this $\mathbb{Z}_4$ symmetry can be understood as follows. Under KW duality, a $\mathbb{Z}_2^x$ domain wall is mapped to a dual $\mathbb{Z}_2^x$ charge. As illustrated in FIG.~\ref{fig:type-II anomaly}(c), the mixed anomaly further implies that such a dual $\mathbb{Z}_2^x$ charge is bound with a half $\mathbb{Z}_2^a$ charge,  {which exactly corresponds to the unit charge of the $\mathbb{Z}_4$ symmetry}. Furthermore, the $\mathbb{Z}_2^a$-SSB phase is correspondingly mapped to the $\mathbb{Z}_4$-SSB phase. Indeed, the dual system must spontaneously break the dual $\mathbb{Z}_2^x$ symmetry (that is dual to the original unbroken $K=\mathbb{Z}_2^x$), which is the only nontrivial subgroup of $\mathbb{Z}^r_4$. Thus, the full $\mathbb{Z}^r_4$ symmetry is broken. Since the $S^1$-families of the $\mathbb{Z}^r_4$-SSB phase are classified by $\mathbb{Z}_4$, the $S^1$-families of the original $\mathbb{Z}_2^a$-SSB phase inherit the same $\mathbb{Z}_4$ classification through the one-to-one correspondence by KW duality.

This $\mathbb{Z}_4$ structure can also be understood directly in the original variables. The mixed anomaly implies that a $\mathbb{Z}_2^a$-domain-wall carries a half $\mathbb{Z}_2^x$-charge, as shown in FIG.~\ref{fig:type-II anomaly}(c).
Thus, an evolution that pumps two $\mathbb{Z}_2^a$-domain-walls necessarily pumps one $\mathbb{Z}_2^x$-charge. Such an evolution is therefore noncontractible and realizes the order-two element of the $\mathbb{Z}_4$ classification.

{We now explicitly construct the noncontractible
loops, where the family of Hamiltonian $
H(\theta)$ is generated by the  unitary evolution operator $\mathcal V(\theta)$.} 

We start with the dual system, in which a  noncontractible loop of $\Z^r_4$-SSB phase is generated by the unitary operator analogously to Eq.~\eqref{kw_0}:
\begin{equation}
    \tilde{\mathcal{V}}_{\Z^{r}_4\text{-SSB}}(\theta)=\exp\left(\frac{ i\theta}{8}\sum^L_{j=1}(1-\tilde{\sigma}^x_j)\right).
\end{equation}
At $\theta=2\pi$, it reduces to $\tilde{U}_r$  and thus pumps a $\Z^r_4$-domain-wall.
Then the corresponding operator before KW duality in the original variables is 
\begin{equation}
    \mathcal{V}_{\Z^a_2\text{-DW}}(\theta)=\exp\left(\frac{i\theta }{8}\sum^L_{j=1}(1-\sigma^z_{j-1}\sigma^z_{j})\right),
\end{equation}
which is $\Z^a_2\times\Z^x_2$ symmetric. 
Since $\mathcal{V}_{\Z^a_2\text{-DW}}(2\pi)=U_a$, it commutes with any $\mathbb{Z}_2^a\times\mathbb{Z}_2^x$-symmetric Hamiltonian, and the corresponding loop pumps a $\mathbb{Z}_2^a$ domain wall. Moreover, due to $\mathcal{V}_{\Z^a_2\text{-DW}}(2\theta)=\mathcal{V}_{\text{charge}} (\theta)$, we conclude that the noncontractible loop pumping two $\mathbb{Z}^a_2$ domain walls, generated by $\mathcal{V}_{\Z^a_2\text{-DW}}(2\theta)$, necessarily pumps one $\mathbb{Z}^x_2$ charge. 
This precisely reproduces the result obtained by the anomaly argument above.

The above analysis can be similarly applied to the $S^1$-families of gapped phases with other SSB structures with anomalous $\mathbb{Z}_2\times\mathbb{Z}_2$ symmetry. Furthermore, it provides a framework for studying $S^1$-families of gapped phases with $(\mathbb{Z}_2)^3$ symmetry that exhibit mixed anomalies. The corresponding classification~\cite{suppl} is summarized in Table~\ref{table1}. 

\parR{$S^1$-family of $\Z_2$-SSB phases with $D_{2n}$ symmetry and Rep$(D_{2n})$-NISPT  phases}
We now turn to a different type of interplay between the unbroken and broken symmetry sectors, arising from a semidirect-product structure.
As a representative example, consider a $D_{2n}=\mathbb Z_n^r\rtimes\mathbb Z_2^s$ symmetric system in a $\mathbb Z_2^s$-SSB phase, whose twofold-degenerate ground states break the $\mathbb Z_2^s$ subgroup.
As we show below, the broken $\mathbb{Z}_2^s$ sector and the unbroken subgroup are intertwined by the semidirect-product structure, which constrains the algebraic structure of the corresponding $S^1$ families. Interestingly, it was shown that $\operatorname{Rep}(D_{2n})$ symmetry can be obtained from $D_{2n}$ symmetry by gauging the $\mathbb Z_2^s$ subgroup, under which the  Rep$(D_{2n})$ NISPT phases are mapped to $\Z^s_2$-SSB phases \cite{4zfz-x9xh}. Therefore the $S^1$ families of $\operatorname{Rep}(D_{2n})$ NISPT phases are in one-to-one correspondence with those of the dual $\mathbb Z_2^s$-SSB phases. As we show below, the resulting classification separates into two cases depending on whether \(n\) is even or odd.

\begin{enumerate}
    \item For odd $n$, the unbroken group of the $\Z^s_2$-SSB phase is $K=\Z^r_n$ and the dual system realizes the trivial NISPT phase \footnote{This NISPT phase is considered trivial, as it can be realized by a product-state ground state on the lattice without edge modes, as demonstrated in Ref.~\cite{4zfz-x9xh} and Eq.~\eqref{eq:Trivial NISPT Hal}.}. 
    The $S^1$-family is generated by loops pumping either $\Z^s_2$-domain wall or $\Z^r_n$-charge, which however do not commute. Since the two SSB ground states are exchanged by $\Z^s_2$, the pumped $\Z^r_n$  charges associated with them must be inverse to each other, as dictated by the group relation  $rs=sr^{-1}$. It is therefore sufficient to track the pumped $\Z^{r}_n$-charge in a single reference ground state. From the combined evolutions illustrated in FIG.~\ref{fig:type-II anomaly}(d), we find that the domain-wall-pumping loop conjugates the charge-pumping loop to its inverse, which implies the $S^1$-family of $\Z^s_2$-SSB is classified by $D_{2n}$ group. By duality, the same classification applies to the trivial Rep$(D_{2n})$-NISPT phase. 

    
    \item For even $n=2m$, two distinct SSB patterns are possible, with the unbroken group $K=\Z^r_n$ or $K=\Z^{r^2}_{m}\rtimes\Z^{rs}_2$. In the first branch, the same argument as above applies and the $S^1$-families of both the $\Z^s_2$-SSB phase and its dual trivial Rep$(D_{2n})$-NISPT phase  are again classified by $D_{2n}$. 
    
    In the second branch, the unbroken group is $D_{2m}=\mathbb Z_{m}^{r^2}\rtimes\mathbb Z_2^{rs}=\{r^2|r^n=1\}\rtimes \{1,rs\}$ and the dual system realizes nontrivial NISPT phases \cite{4zfz-x9xh}. Following an analysis~\cite{suppl} analogous to that above, we obtain that $S^1$-family classification is $D_8$  for even $m$ while it is $\Z_2\times\Z_2$ for odd $m$.

     \end{enumerate}
We remark that unlike conventional SPT phases with group symmetry, the $S^1$-family classification of NISPT phases can depend on whether the underlying NISPT phase where the $S^1$ (or its base point) stay is trivial or nontrivial. For conventional SPT phases, the $S^1$-family classification is independent of the base point position \cite{Kitaev2011SRE,Kitaev2013SRE,PhysRevB.110.094410}. This distinction can be traced to the invertible stacking structure of conventional SPT phases, which allows different SPT phases to be related by stacking with another SPT phase. By contrast, NISPT phases are intrinsically noninvertible, and their $S^1$-family structure can therefore depend on the specific NISPT phase hosting the $S^1$.  {Interestingly, for $n=4$, both the trivial and nontrivial NISPT phases have an $S^1$-family classified by $D_8$, consistent with the result obtained from autoequivalences of the corresponding module categories of the noninvertible symmetry~\cite{li2025classification}. }

Now let us construct loops of the first branch, whose $S^1$-family classification is $D_{2n}$, in a qudit chain.
On each vertex, we introduce a qubit $\ket{s^\tau_j}, s^\tau_j\in\mathbb Z_2$ and a $\mathbb Z_n$ qudit $\ket{s_j}, s_j\in\mathbb Z_n$.
The clock and shift operators acting on a vertex $j$ are
\begin{equation}
Z_j=\sum_{s=0}^{n-1}e^{\frac{2\pi i}{n}s}\ket{s}\bra{s}_j, \ X_j=\sum_{s=0}^{n-1}\ket{s+1}\bra{s}_j.
\end{equation}
The $D_{2n}$ symmetry is generated by
\begin{equation}\label{eq:lattice D8 sym}
    U_r=\prod_{j} X_j,\quad  U_s=\prod_{j} C_j \tau^x_j, 
\end{equation}
where $C_j=\sum_{s=0}^{n-1}\ket{-s}\bra{s}_j$ is the charge conjugation on vertex $j$.  We consider the $\Z^s_2$-SSB Hamiltonian
\begin{equation}\label{eq: Z2SSB main text}
H_{\Z^s_2\text{-SSB}}=-\sum_j X_j+X^{\dagger}_j-\sum_j \tau^z_j \tau^z_{j+1}.
\end{equation}
Then the noncontractible loops, i.e., the family of Hamiltonian $
H(\theta)=\mathcal V(\theta)H_{\Z^s_2\text{-SSB}}\mathcal V^\dagger(\theta)$, can be generated by
\begin{equation}\label{eq:D2n evolution loop}
\begin{split}
&\mathcal{V}_{\Z^r_n\text{-charge}}(\theta)=\prod_j\exp\left(\frac{\theta}{2\pi}\text{ln}[Z^{-\tau^z_{j+1}}_{j}Z^{\tau^z_{j+1}}_{j+1}]\right),\\ &\mathcal{V}_{\Z^s_2\text{-DW}}(\theta)=\prod_j\exp\left(\frac{i\theta}{4}(1-\tau^x_j)\right).
\end{split}
\end{equation}
The first  loop pumps the $\pm 1$ of $\Z^r_n$ charge in the  ground states with $\tau^z=\pm 1$, respectively, and the second loop pumps the $\Z^s_2$ domain wall; both loops are $D_{2n}$ symmetric.

To gauge $\Z^s_2$, we introduce the  $\mathbb{Z}_2$ gauge field $\mu_{j+\frac{1}{2}}$, i.e., a qubit, on each link.  
Following the gauging procedure in Refs.~\cite{SciPostPhys.17.4.115,4zfz-x9xh},
we obtain the dual loops generated by
\begin{equation}
\begin{split}
&\mathcal V_{\gamma_1}(\theta)=\prod_j\exp\left(\frac{\theta}{2\pi}\text{ln}[Z^{-\mu^x_{j+\frac{1}{2}}}_{j}Z_{j+1}]\right),\\ 
&\mathcal V_{\gamma_2}(\theta)=\prod_j\exp\left(\frac{i\theta}{4}(1-C_j\mu^z_{j-\frac{1}{2}}\mu^z_{j+\frac{1}{2}})\right),
\end{split}
\end{equation}
together with the dual Hamiltonian
\begin{equation}\label{eq:Trivial NISPT Hal}
H_{\text{NISPT}}=-\sum_j X_j+X^{\dagger}_j-\sum_j \mu^x_{j+\frac{1}{2}},
\end{equation}
reproducing the $D_{2n}$ group-structure of $S^1$-family~\cite{Supple}.
\parR{Conclusion}
In this work,
we identify the $S^1$-families of the gapped quantum phase diagram of systems,
where the unbroken and spontaneously broken symmetries exhibit mixed anomalies or semidirect-product structures, as well as those of the NISPT phase diagram. We show that these exotic symmetries give rise to novel structures of the $S^1$-families. We further construct lattice realizations of the noncontractible loops in these $S^1$-families on tensor-product Hilbert spaces, 
which directly reveal their associated algebraic structures.
Our framework can be further applied to the classification and lattice construction of $S^1$-families with more general noninvertible symmetries, such as $G\times \mathrm{Rep}(G)$, as well as to higher-dimensional systems, 
which we leave for future work.

\parR{Acknowledgments.}
The authors thank Chenjie Wang for useful discussions.
 L.H.L. is supported by a Quantum SuperSEED fund and a startup fund from the Pennsylvania State University (Zhen Bi).
The work of Y. Y. was supported by the National Key Research and Development Program of China (Grant No.~2024YFA1408303), the National Natural Science Foundation of China (Grants No.~12474157 and No.~12447103), the sponsorship from Yangyang Development Fund, and Xiaomi Young Scholars Program.
\bibliography{ref}

\newpage
\onecolumngrid
\appendix
\newpage
\section{Group structure of $S^1$-family}
We study the $S^1$-family of gapped phases represented by the  Hamiltonians $H(\theta)$, parameterized by $ 0\le\theta \le 2\pi$, with the periodicity condition $H(2\pi)=H(0)$. Two such loops $\gamma_1$ and $\gamma_2$ are regarded in the same class if they can be smoothly connected by a continuous family of gapped Hamiltonians $H(\theta, r)$ satisfying $H(\theta,r=0)=H_{\gamma_1}(\theta)$ and $H(\theta,r=1)=H_{\gamma_2}(\theta)$. We denote the equivalence class of a loop $\gamma$ as $[\gamma]$.
The set of equivalence classes naturally admits an additive structure under loop concatenation. For example,
\begin{equation}
    H_{\gamma_3}(\theta)=\begin{cases}
H_{\gamma_1}(2\theta), & 0\le\theta<\pi,\\
H_{\gamma_2}(2\theta-2\pi), & \pi\le \theta<2\pi,
\end{cases}
\end{equation}
realizes a loop in  equivalence classes $[\gamma_1+\gamma_2]$.

In the following, we construct symmetric unitary evolution operators $\mathcal V(\theta)$ with $\theta\in[0,2\pi]$ that generate noncontractible loops of Hamiltonian according to $
H(\theta)=\mathcal V(\theta)H(0)\mathcal V^\dagger(\theta)$, where $H(0)$ is a Hamiltonian realizing the corresponding phase, and the periodicity condition $H(0)=H(2\pi)$ requires that $\mathcal V(2\pi)$ commutes with $H(0)$.
\\
~\\
~\\

\section{$S^1$-family of gapped phases with anomalous $\Z_2\times \Z_2$ symmetry in 1D}
In this section, we study the  $S^1$-family of all possible gapped phases with $\Z_2\times \Z_2$ symmetry in 1D, which carries the mixed anomaly.

\subsection{$S^1$-family of $\Z_2$-SSB phases}
We first consider two SSB phases, in each of which only one $\Z_2$ is spontaneously broken. Due to the mixed anomaly, a domain wall of one $\Z_2$ symmetry must carry a half charge of the other $\Z_2$ symmetry. It then follows that the $S^1$-family of   each of  the two $\Z_2$-SSB phases is classified by $\Z_4$.

We now employ the Kramers–Wannier duality to construct a lattice realization of these noncontractible loops. To this end, we introduce two qubits on each site, with Pauli matrices
$\sigma^{x,y,z}_{j}$ and $\mu^{x,y,z}_{j}$ and denote the action of $\Z^a_2\times\Z^x_2$ on the enlarged Hilbert space as:
\begin{equation}
 U'_a =\prod_j\exp\left(\frac{\pi i}{4} (1-\sigma^z_{j-1}\sigma^z_j)\right)\mu^x_j,  U_x=\prod_j\sigma^x_j,
\end{equation}
where $U_x$ acts trivially on the $\mu$ qubits. We note that $U'_a$ acts in the same way as $U_a$ in the main text on the $\sigma$-spins and therefore exhibits the same mixed anomaly with $U_x$.  This choice of lattice realization is made solely for the convenience of constructing the noncontractible loops.

Now let us consider the $\Z^x_2$-SSB phase since the discussion of the $\Z^a_2$-SSB phase is the same as that in the main text.

The $\Z^x_2$-SSB  Hamiltonian is
\begin{equation}
H_{\Z^x_2 \text{-SSB}}=-\sum_j \sigma^z_j\sigma^z_{j+1}-\sum_j \mu^x_{j}.
\end{equation}

Under the KW duality for $\sigma$-spins, the  $\Z^a_2\times\Z^x_2$ symmetry is mapped to $\Z^r_4$ symmetry generated by
\begin{equation}
 U'_r =\prod_j\exp\left(\frac{\pi i}{4} (1-\tilde{\sigma}^x_j)\right)\mu^x_j
\end{equation}
and the $\Z^x_2$-SSB  Hamiltonian is mapped to a $\Z_4$ symmetric phase Hamiltonian 
\begin{equation}
H_{\Z^r_4 \text{-sym}}=-\sum_j \tilde{\sigma}^x_j-\sum_j \mu^x_{j}.
\end{equation}
Since the $S^1$-family classification of $\Z_4$ symmetric phase is $\Z_4$, the corresponding classification of $\Z^x_2$-SSB phase is also $\Z_4$.

We can then construct the evolution loop that pumps $\Z^r_4$ charge:
\begin{equation}
    \tilde{\mathcal{V}}_{\Z^r_4\text{-charge}}(\theta)=\prod_j\exp\left(\frac{\theta}{2\pi}\text{ln}[\tilde{\sigma}^z_j \text{CNOT}_{j}\tilde{\sigma}^z_{j+1}\mu^z_{j+1} \text{CNOT}_{j+1}]\right),
\end{equation}
where $\text{CNOT}_{j}$ is the Controlled-NOT gate from $\sigma$ to $\mu$ given by:
\begin{equation}
\text{CNOT}_{j}=\frac{1+\tilde{\sigma}^x_j}{2}+\mu^z_j\frac{1-\tilde{\sigma}^x_j}{2}.
\end{equation}
To further explain the physics of this evolution loop, we note the Hilbert space of the two qubits on each site can be mapped to that of a $\Z_4$ qudit. We consider the basis state $|s^{\mu}, s^{\tilde{\sigma}}\rangle$ with eigenvalues $((-1)^{s^{\mu}},(-1)^{s^{\tilde{\sigma}}})$ of operators $(\mu^x,\sigma^x)$ and $s^{\mu},s^{\tilde{\sigma}}=0,1$. Then the $\Z_4$ qudit state is labeled by 
\begin{equation}\label{eq:Z4}
s=s^{\tilde{\sigma}}+2s^{\mu}\in\Z_4.
\end{equation}
We then define the generalized Pauli operators
\begin{equation}
X_j=\sum^{3}_{s_j=0}i^{s_j}|s_j\rangle\langle s_j|, Z_j=\sum^{3}_{s_j=0}|s_j-1\rangle\langle s_j|, iX_j Z_j=Z_j X_j.
\end{equation}
In this basis, the $\Z^r_4$ symmetry operator takes the form
\begin{equation}
U'_r =\prod_j X_j,
\end{equation}
and the above evolution loop becomes
\begin{equation}
 \tilde{\mathcal{V}}_{\Z^r_4\text{-charge}}(\theta)=\prod_j\exp\left(\frac{\theta}{2\pi}\text{ln}[Z^{\dagger}_jZ_{j+1}]\right).
\end{equation}
When $\theta=2\pi$, truncating this evolution loop to the region $  l\le j\le k$ yields
\begin{equation}
    \mathcal{V}^{\text{TR}}_{\Z^r_4\text{-charge}}(2\pi)=Z^{\dagger}_{l}Z_{k},
\end{equation}
which carries the $\Z^r_4$ charge on each boundary.

Applying the inverse KW duality, this evolution is mapped to
\begin{equation}\label{eq:loop bdw}
    \mathcal{V}_{\Z^x_2\text{-DW}}(\theta)=\prod_j\exp\left(\frac{\theta}{2\pi}\text{ln}[\sigma^x_j \text{CNOT}'_{j}\mu^z_{j+1} \text{CNOT}'_{j+1}]\right)=\prod_j\exp\left(\frac{\theta}{2\pi}\text{ln}[ \text{CNOT}'_{j}\sigma^x_j \mu^z_j\mu^z_{j+1}\text{CNOT}'_{j+1}]\right),
\end{equation}
where
\begin{equation}
\text{CNOT}'_{j}=\frac{1+\sigma^z_{j-1}\sigma^z_j}{2}+\mu^z_j\frac{1-\sigma^z_{j-1}\sigma^z_j}{2}.
\end{equation}
This evolution is $\Z^a_2\times\Z^x_2$ symmetric throughout the whole loop. Under periodic boundary conditions (PBC), we have
\begin{equation}
 \mathcal{V}_{\Z^x_2\text{-DW}}(2\pi)=\prod_j \sigma^x_j,
\end{equation}
which pumps the $\Z^x_2$ domain wall. Moreover, when $\theta=4\pi$, truncating the evolution loop to the region $ l\le j\le k$ gives
\begin{equation}\label{eq:truncated loop bdw}
    \mathcal{V}^{\text{TR}}_{\Z^x_2\text{-DW}}(4\pi)=\prod_{l\le j\le k-1}[\sigma^x_j \text{CNOT}'_{j}\mu^z_{j+1} \text{CNOT}'_{j+1}]^2=\prod_{l\le j\le k-1}\mu^z_{j}\mu^z_{j+1}=\mu^z_l\mu^z_{k}.
\end{equation}
Each boundary operator carries a $\Z^a_2$ charge,  consistent with the expected $\Z_4$ classification.


\subsection{$S^1$-family of fully SSB phases}
When $\Z^a_2\times \Z^x_2$ are both broken, the $S^1$-family classification is simply $\Z_2\times \Z_2$ since the two types of domain walls are independent. The corresponding noncontractible loops are generated by the evolutions
\begin{equation}
    \mathcal{V}_{\Z^a_2\text{-DW}}(\theta)=\exp\left(\frac{i\theta }{8}\sum^L_{j=1}(1-\sigma^z_{j-1}\sigma^z_{j})\right)\exp\left(\frac{i\theta}{4}\sum^L_{j=1}(1-\mu^x_{j})\right),\quad \mathcal{V}_{\Z^x_2\text{-DW}}(\theta)=\exp\left(\frac{i\theta }{4}\sum^L_{j=1}(1-\sigma^x_{j})\right).
\end{equation}
This result can also be understood from the dual perspective, where the dual system is in an SSB phase breaking $\Z^r_4$ to the normal $\Z_2$ generated by $\prod_j\tilde{\sigma}^x_j$. Thus the corresponding $S^1$-family
then arises from pumping the domain wall of the broken $\Z_2$ and the charge of unbroken $\Z_2$. These are generated by the evolutions
\begin{equation}
    \tilde{\mathcal{V}}_{\text{DW}}(\theta)=\exp\left(\frac{i\theta}{8}\sum^L_{j=1}(1-\tilde{\sigma}^x_{j})\right)\exp\left(\frac{i\theta}{4}\sum^L_{j=1}(1-\mu^x_{j})\right),\quad  \tilde{\mathcal{V}}_{\text{charge}}(\theta)=\exp\left(\frac{i\theta}{4}\sum^L_{j=1}(1-\tilde{\sigma}^z_{j-1}\tilde{\sigma}^z_{j})\right),
\end{equation}
which are exactly KW dual of $ \mathcal{V}_{\Z^a_2\text{-DW}}$ and $ \mathcal{V}_{\Z^x_2\text{-DW}}$ respectively.
\\
~\\
~\\

\section{$S^1$-family of gapped phases with anomalous $\Z_2\times \Z_2\times\Z_2$ symmetry}
In this section, we study the $S^1$-family of gapped phases with $\Z_2\times \Z_2\times\Z_2$ symmetry carrying a type-III mixed anomaly. Physically, this anomaly implies that one $\Z_2$ is a duality connecting the trivial and nontrivial symmetry-protected topological (SPT) phases protected by the remaining $\Z_2\times \Z_2$ symmetry. A lattice realization is \cite{PhysRevB.106.224420}
\begin{equation}\label{eq: type3 anomaly ope}
    U_o=\prod_j \sigma^x_{2j-1},  U_e=\prod_j \sigma^x_{2j}, U_a =\exp\left(\frac{\pi i}{4} \sum_j (1-\sigma^z_{j-1}\sigma^z_j)\right).
\end{equation}
Here $U_a$ connects trivial and nontrivial $\Z^o_2\times \Z^e_2$-SPT phases, realized by $H=-\sum_j \sigma^x_j$ and   $H=\sum_j \sigma^z_{j-1}\sigma^x_j\sigma^z_{j+1}$  respectively.

\subsection{$S^1$-family of $\Z_2$-SSB phases}
Let us first consider the SSB phase where only one $\Z_2$ is broken. We take the $\Z^{o}_2$-SSB phase with unbroken symmetry  $K=\Z^e_2\times\Z^a_2$ as an example, which can be realized by 
\begin{equation}
H_{\Z^o_2\text{-SSB}}=\sum_{j}(\sigma^x_{2j}-\sigma^z_{2j-1}\sigma^x_{2j}\sigma^z_{2j+1}).
\end{equation}
Under the KW duality, the symmetry is mapped to the dihedral group $D_8=\Z^r_4\rtimes\Z^s_2$, generated by
\begin{equation}
      \tilde{U}_r= \exp\left(\frac{\pi i}{4} \sum_j (1-\tilde{\sigma}^x_{j})\right), \tilde{U}_s =\prod_j \tilde{\sigma}^z_j.
\end{equation}
The dual Hamiltonian is given by
\begin{equation}
\tilde{H}_{D_8\text{-sym}}=\sum_{j}(\tilde{\sigma}^z_{2j}\tilde{\sigma}^z_{2j+1}+\tilde{\sigma}^y_{2j}\tilde{\sigma}^y_{2j+1}),
\end{equation}
which has a unique gapped ground state and realizes a $D_8$ symmetric phase.

This statement can be generalized to any SSB phase breaking a single $\Z_2$ symmetry in $(\Z_2)^3$. Gauging the broken $\mathbb{Z}_2$ always maps the system to a dual system with $D_8$ symmetry, where the original $\mathbb{Z}_2$-SSB phase is dual to a $D_8$ symmetric phase. Since the $S^1$-family classification of the $D_8$ symmetric phase is $H^1(D_8, U(1))=\mathbb{Z}_2\times \mathbb{Z}_2$, the corresponding $S^1$-family classification of such $\mathbb{Z}_2$-SSB phases is likewise $\mathbb{Z}_2\times \mathbb{Z}_2$.


This result  can also be understood directly from the mixed anomaly. Let us again consider $\Z^o_2$ SSB as an example. As discussed above, the candidate noncontractible evolutions pump either a $\Z^o_2$ domain wall or a charge of the unbroken group. However, the mixed anomaly implies that the $\Z^o_2$ domain wall carries a projective representation of $\Z^e_2\times\Z^a_2$. In particular, in the presence of a  $\Z^o_2$ domain wall, i.e. the symmetry-twisted boundary condition $\sigma^z_{L+1}=U_o \sigma^z_1 U^{\dagger}_o=-\sigma^z_1$, the symmetry operator $U_a$ is modified as
\begin{equation}
    U^{\text{mod}}_{a} =\exp\left(\frac{\pi i}{4} \sum^L_{j=2} (1-\sigma^z_{j-1}\sigma^z_j)\right)\exp\left(\frac{\pi i}{4} (1+\sigma^z_1\sigma^z_L)\right),
\end{equation}
and the others are invariant. This modified $U_a$ anticommutes with $U_e$, rendering the $\Z^e_2\times \Z^a_2$-charge ill-defined \cite{PhysRevB.106.224420}. As a result, for a unitary evolution loop  which pumps $\Z^o_2$ domain wall, the corresponding pumped $\Z^e_2\times \Z^a_2$-charge is ill-defined. Thus this evolution loop cannot preserve both the $\Z^e_2\times \Z^a_2$ symmetry. On the other hand, the evolution is required to preserve the full $(\Z_2)^3$ symmetry throughout the entire loop. This contradiction implies that no $(\Z_2)^3$ symmetric evolution can pump a $\Z^o_2$ domain wall.  For example, a loop pumping $\Z^o_2$ domain wall 
\begin{equation}
    \mathcal{V}_{\Z^o_2\text{-DW}}(\theta)=\exp\left(\frac{i\theta }{4}\sum^L_{j=1}(1-\sigma^x_{2j-1})\right)
\end{equation}
doesnot commute with $\Z^a_2$ symmetry. This argument applies to any $\Z_2$-SSB phase. Therefore the corresponding $\Z_2\times \Z_2$ $S^1$-family classification comes from evolutions pumping charges of the unbroken $\Z_2\times\Z_2$ subgroup. 

On the lattice, we construct the nontrivial evolution loops of $\Z^a_2$-SSB phase:
\begin{equation}
    \mathcal{V}_{\Z^o_2\text{-charge}}(\theta)=\exp\left(\frac{i\theta }{4}\sum^L_{j=1}(1-\sigma^z_{2j-1}\sigma^z_{2j+1})\right), \mathcal{V}_{\Z^e_2\text{-charge}}(\theta)=\exp\left(\frac{i\theta }{4}\sum^L_{j=1}(1-\sigma^z_{2j}\sigma^z_{2j+2})\right),
\end{equation}
which are $(\Z_2)^3$ symmetric and satisfy that $\mathcal{V}_{\Z^o_2\text{-charge}}(2\pi)=\mathcal{V}_{\Z^e_2\text{-charge}}(2\pi)=\mathbb{I}$ under PBC.

\subsection{$S^1$-family of $\Z_2\times \Z_2$-SSB phases}
Next, we study the $S^1$-family of $\Z_2\times \Z_2$-SSB phase. We take  $\Z^o_2\times\Z^e_2$-SSB with unbroken group $K=\Z^a_2$ as an example, which is realized by the Hamiltonian 
\begin{equation}
    H_{\Z^o_2\times \Z^e_2\text{-SSB}}=-\sum_j (\sigma^z_{2j-1}\sigma^z_{2j+1}+\sigma^z_{2j}\sigma^z_{2j+2}).
\end{equation}
The KW duality maps this Hamiltonian to
\begin{equation}
    \tilde{H}_{\Z^s_2 \text{-SSB}}=-\sum_j (\tilde{\sigma}^x_{2j}\tilde{\sigma}^x_{2j+1}+\tilde{\sigma}^x_{2j+1}\tilde{\sigma}^x_{2j+2}),
\end{equation}
which realizes the $\Z^s_2$-SSB phase with unbroken group $K=\Z^{r}_4$.
According to the main text, the $S^1$-family of such $\Z^s_2$-SSB is classified by $D_8$ group. By duality, the same classification applies to the $\Z^o_2\times \Z^e_2$-SSB phase. We will give the lattice realization of the corresponding noncontractible loops at the end of SM.

Now let us provide an anomaly-based argument on the above $S^1$-family, through which this classification can be understood directly from the algebraic relations between domain-wall and charge-pumping evolutions of the $\Z^o_2\times\Z^e_2$-SSB phase. 

\begin{figure}[htb]
 \centering
\includegraphics[scale=0.7]{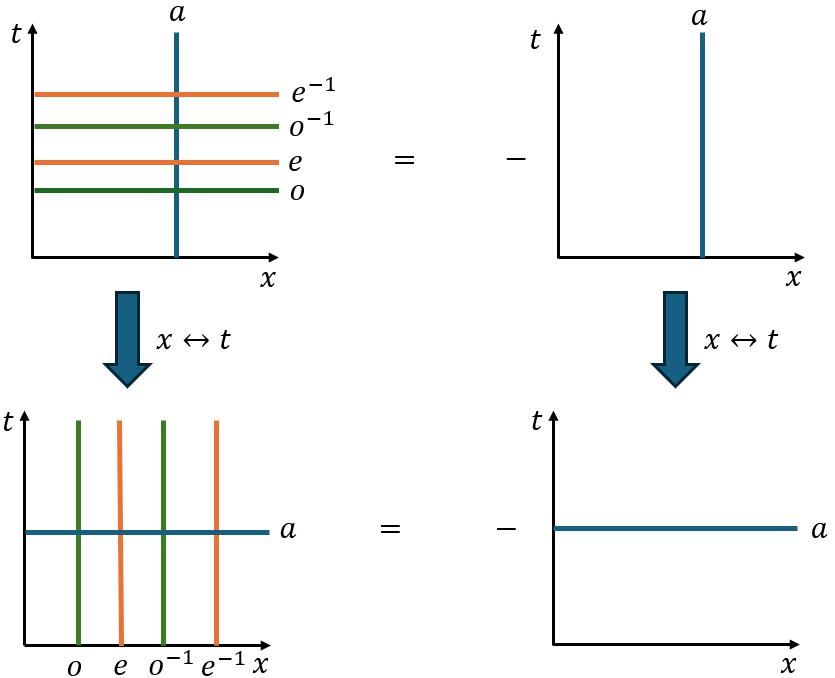}

\caption{The symmetry defect configuration in the spacetime. Here the symmetry defect lines which are vertical to $x$ or $t$ direction correspond to inserting the domain walls or adding symmetry operators in the path integral.} 

\label{fig:anomaly}
\end{figure}

We first analyze the path integral in the symmetry-defect configuration shown in FIG. \ref{fig:anomaly}. The upper-left and upper-right configurations  imply
\begin{equation}
\text{Tr}(\tilde{U}_o \tilde{U}_e \tilde{U}^{-1}_o  \tilde{U}^{-1}_e e^{iH_{a\text{-DW}}t})=-\text{Tr}( e^{iH_{a\text{-DW}}t}).
\end{equation}
Here $\tilde{U}_o$ and $\tilde{U}_e$ denote the modified  $\Z^o_2\times\Z^e_2$ symmetry operators in the presence of a $\Z^a_2$ domain wall. This relation comes from the type-III mixed anomaly \cite{PhysRevLett.114.031601,PhysRevB.91.035134} and shows that the modified symmetry operators anticommute in the presence of the domain wall.  Exchanging $x$ and $t$, we interchange  the roles of domain walls and symmetry operators. Under this perspective, the lower-left configuration corresponds to inserting, from left to right, a $\Z^o_2$ domain wall, a $\Z^e_2$ domain wall, the inverse of a $\Z^o_2$ domain wall and the inverse of a $\Z^e_2$ domain wall. The relation between  the lower-left and lower-right configurations in FIG.~\ref{fig:anomaly} then implies that this domain-wall configuration carries a nontrivial $\Z^a_2$-charge. 

Then we consider evolutions $\mathcal{V}_{\Z^o_2\text{-DW}}, \mathcal{V}_{\Z^e_2\text{-DW}}$ and $\mathcal{V}_{\Z^a_2\text{-charge}}$, which pump the broken $\Z^o_2\times \Z^e_2$ domain wall and $\Z^a_2$ charge, respectively. To reveal the algebraic structure, we consider a combined evolution loop consisting of the $\mathcal{V}_{\Z^o_2\text{-DW}}$ loop, followed by the $\mathcal{V}_{\Z^e_2\text{-DW}}$ loop, then by the inverse of $\mathcal{V}_{\Z^o_2\text{-DW}}$  and finally the inverse of the $\mathcal{V}_{\Z^e_2\text{-DW}}$  loop.
This realizes the homotopy class $[\mathcal{V}_{\Z^o_2\text{-DW}}]+[\mathcal{V}_{\Z^e_2\text{-DW}}]-[\mathcal{V}_{\Z^o_2\text{-DW}}]-[\mathcal{V}_{\Z^e_2\text{-DW}}]$. This combined evolution precisely corresponds to the domain wall configuration in the lower-left figure in FIG.~\ref{fig:anomaly}. Then by the relation in the last paragraph, it must pump a $\Z^a_2$ charge. This nontrivial algebraic relation leads directly to the $D_8$ classification.

Here we remark that this argument does not depend on the particular choice of broken $(\Z_2)^2$ symmetry and therefore applies generally to any $(\Z_2\times \Z_2)$-SSB  phase.

\subsection{$S^1$-family of fully SSB phase}

Finally, we consider the $S^1$-family of SSB phases in which the entire $(\Z_2)^3$ symmetry is broken. The corresponding nontrivial evolution loops pump the three independent domain walls, leading directly to the $(\Z_2)^3$ classification.
\\
~\\
~\\

\section{$S^1$-family of Rep$(D_{2n})$-NISPT  phases}
In this section, we discuss $S^1$-family of Rep$(D_{2n})$ noninvertible SPT (NISPT) phases. Reference \cite{4zfz-x9xh} proved that Rep$(D_{2n})$ symmetry can be obtained from $D_{2n}=\Z^{r}_n\rtimes \Z^s_2$ by gauging $\Z^s_2$ symmetry, under which the  NISPT phases are dual to $\Z^s_2$-SSB phases. Therefore the $S^1$-family of Rep$(D_{2n})$-NISPT phases is the same as that of the corresponding $\Z^s_2$-SSB phases. 

\subsection{Classification of $S^1$-families of Rep$(D_{2n})$-NISPT  phases}
We focus on the second branch for even $n=2m$ since the first branch has been discussed in the main text.
    
In the second branch for even $n=2m$, the unbroken group of $\Z^s_2$-SSB is $D_{2m}=\mathbb Z_{m}^{r^2}\rtimes\mathbb Z_2^{rs}=\{r^2|r^n=1\}\rtimes \{1,rs\}$ and its charges, i.e. 1-dimensional representations, are classified by $\Z_2\times\Z_2$ for even $m$ while $\Z_2$ for odd $m$.

For even $m$, there are two such $\Z^s_2$-SSB phases, which differ by stacking with a nontrivial SPT phase protected by the unbroken $D_{2m}$ symmetry group \cite{4zfz-x9xh}. However, they have the same $S^1$-family, since distinct SPT phases with the same group symmetry share the same $S^1$-family associated with charge pumping \cite{PhysRevB.110.094410}. In this case, the $\Z_2\times\Z_2$ classification of unbroken symmetry charges corresponds to pumping either a $\Z_2$-charge of the $\Z^{rs}_2$ or $\Z^{r^2}_{m}$ symmetry.  Since the two SSB ground states are connected by $\Z^s_2$, if one ground state carries with nontrivial $\Z_2$-charge of $\Z^{r^2}_{m}$ symmetry, the other must carry nontrivial $\Z_2$-charges of both the $\Z^{rs}_2$ and $\Z^{r^2}_{m}$ symmetry, due to $s(rs)s^{-1}=r^{-2}(rs)$. It is therefore sufficient to track the pumped charge in a single reference ground state. Therefore, by conjugating the unitary loop which pumps the $\Z^s_2$ domain wall, the unitary loop which only pumps the $\Z_2$-charge of the $\Z^{r^2}_{m}$ symmetry is mapped to a unitary loop which pumps both $\Z_2$-charges of the $\Z^{rs}_2$ and $\Z^{r^2}_{m}$ symmetry. Together with the $\Z^s_2$
 domain-wall pumping evolution, this relation implies that $S^1$-family classification is $D_8$.
 
 For odd $m$, the only nontrivial pumped charge comes from the \(\mathbb{Z}_2^{rs}\) symmetry. Combining this with the $\Z^s_2$
 domain-wall pumping evolution, one finds that $S^1$-family classification is $\Z_2\times\Z_2$.


Then we construct the evolution loops generating the $S^1$-family of $\text{Rep}(D_{2n})$-NISPT on the lattice.

\subsection{Evolution loops of trivial $\text{Rep}(D_{2n})$ NISPT}
We start with the NISPT dual to the $\Z^s_2$-SSB phase with unbroken group $K=\Z^r_n$. This NISPT phase is denoted as a trivial NISPT as it can be realized by a product ground state \cite{4zfz-x9xh}.

On each vertex, we assign a qubit $\ket{s^\tau_j}, s^\tau_j\in\mathbb Z_2$ and a $\mathbb Z_n$ qudit $\ket{s_j}, s_j\in\mathbb Z_n$.
The clock and shift operators acting on a vertex $j$ are
\begin{equation}
Z_j=\sum_{s=0}^{n-1}e^{\frac{2\pi i}{n}s}\ket{s}\bra{s}_j, \ X_j=\sum_{s=0}^{n-1}\ket{s+1}\bra{s}_j.
\end{equation}
The $D_{2n}$ symmetry is generated by
\begin{equation}\label{eq:lattice D8 sym}
    U_r=\prod_{j} X_j,\quad  U_s=\prod_{j} C_j \tau^x_j, 
\end{equation}
where $C_j=\sum_{s=0}^{n-1}\ket{-s}\bra{s}_j$ is the charge conjugation on vertex $j$.  The corresponding $\Z^s_2$-SSB Hamiltonian is
\begin{equation}\label{eq:Z2s SSB Hal}
H_{\Z^s_2\text{-SSB}}=-\sum_j X_j+\text{h.c.}-\sum_j \tau^z_j \tau^z_{j+1}.
\end{equation}
Then we construct the evolution loops
\begin{equation}\label{eq:D2n evolution loop}
\mathcal{V}_{\Z^r_n\text{-charge}}(\theta)=\prod_j\exp\left(\frac{\theta}{2\pi}\text{ln}[Z^{-\tau^z_{j+1}}_{j}Z^{\tau^z_{j+1}}_{j+1}]\right),\quad \mathcal{V}_{\Z^s_2\text{-DW}}(\theta)=\prod_j\exp\left(\frac{i\theta}{4}(1-\tau^x_j)\right).
\end{equation}
The first  evolution loop pumps the 1 or $-1$ $\Z^r_n$ charge in the spin-up or spin-down ground state and the second evolution loop pumps the $\Z^s_2$ domain wall, which are both $D_{2n}$ symmetric.

To gauge $\Z^s_2$, we introduce the  $\mathbb{Z}_2$ gauge field $\mu_{j+\frac{1}{2}}$ on each link  and project the extended Hilbert space to the gauge-invariant sector. The Gauss law on each vertex is
\begin{equation}
\begin{split}
G_j=C_j\tau^x_j\mu^z_{j-\frac{1}{2}}\mu^z_{j+\frac{1}{2}}=1. 
 \end{split}
\end{equation}
Minimally coupling the adiabatic evolution loops with gauge field $\mu$, we get the gauge-invariant evolution loops
\begin{equation}
\mathcal{V}^{\text{gauged}}_{\Z^r_n\text{-charge}}(\theta)=\prod_j\exp\left(\frac{\theta}{2\pi}\text{ln}[Z^{-\mu^x_{j+\frac{1}{2}}\tau^z_{j+1}}_{j}Z^{\tau^z_{j+1}}_{j+1}]\right),\quad \mathcal{V}^{\text{gauged}}_{\Z^s_2\text{-DW}}(\theta)=\prod_j\exp\left(\frac{i\theta}{4}(1-\tau^x_j)\right).
\end{equation}

Then by applying the unitary transformation
\begin{equation}
  U'=\prod_{j} \left(\frac{(1+\tau^z_j)}{2}+\frac{(1-\tau^z_j)}{2}C_j\mu^z_{j-\frac{1}{2}}\mu^z_{j+\frac{1}{2}}\right)
\end{equation}
to simplify the Gauss law to $\tau^x_j=1$, the gauged 
 evolution loops become
\begin{equation}
\begin{split}
& U'\mathcal V^{\text{gauged}}_{\Z^r_n\text{-charge}}(\theta)U'^{-1}=\prod_j\exp\left(\frac{\theta}{2\pi}\text{ln}[Z^{-\mu^x_{j+\frac{1}{2}}}_{j}Z_{j+1}]\right),\\ &U'\mathcal V^{\text{gauged}}_{\Z^s_2\text{-DW}}(\theta)U'^{-1}=\prod_j\exp\left(\frac{i\theta}{4}(1-\tau^x_jC_j\mu^z_{j-\frac{1}{2}}\mu^z_{j+\frac{1}{2}})\right).
\end{split}
\end{equation}
Finally, by projecting onto the gauge-invariant sector, we obtain the dual 
evolution loops
\begin{equation}
\mathcal V_{\gamma_1}(\theta)=\prod_j\exp\left(\frac{\theta}{2\pi}\text{ln}[Z^{-\mu^x_{j+\frac{1}{2}}}_{j}Z_{j+1}]\right),\quad \mathcal V_{\gamma_2}(\theta)=\prod_j\exp\left(\frac{i\theta}{4}(1-C_j\mu^z_{j-\frac{1}{2}}\mu^z_{j+\frac{1}{2}})\right).
\end{equation}

According to Ref. \cite{4zfz-x9xh}, the dual Rep$(D_{2n})$ symmetry contains a $\mathbb{Z}_2$ operator $\prod_j \mu^x_{j+1/2}$, as well as the noninvertible operators
\[
\begin{split}
\mathsf{W}_{k}=\text{Tr}\left(\delta_{\alpha_1,\alpha_{L+1}}\prod^L_{j=1}M^{(j)}_{\alpha_j\alpha_{j+1}}\right),\quad
M^{(j)}=\begin{bmatrix}
    X^k_j\frac{1+\mu^x_{j+\frac{1}{2}}}{2}       & X^{k}_j\frac{1-\mu^x_{j+\frac{1}{2}}}{2}  \\
     X^{-k}_j\frac{1-\mu^x_{j+\frac{1}{2}}}{2}        & X^{-k}_j\frac{1+\mu^x_{j+\frac{1}{2}}}{2}
\end{bmatrix}.
\end{split}
\]
For odd $n$, the index takes the values $k=1,\ldots,(n-1)/2$, whereas for even $n$, $k=1,\ldots,n/2-1$. In the latter case, $\mathrm{Rep}(D_{2n})$ also contains an additional $\mathbb{Z}_2$ operator $\prod_j X_j^{n/2}$.

Moreover, the dual Hamiltonian is given by
\begin{equation}
H_{\text{NISPT}}=-\sum_j X_j+\text{h.c.}-\sum_j \mu^x_{j+\frac{1}{2}}.
\end{equation}
For the first evolution loop, since it commutes with $\mu^x_{j+1/2}$ for all links, the ground state of $\mu$ is invariant under the evolution. As a result, the truncated
evolution loop with $\theta=2\pi$ in the region $l\le j\le k$ acts on the ground state effectively as
\begin{equation}
\mathcal V^{\text{TR}}_{\gamma_1}(2\pi)|_{\text{G.S.}}=Z^{\dagger}_l Z_k,
\end{equation}
whose boundary operators generate a $\Z_n$ group. We also have $\mathcal{V}_{\gamma_2}(2\pi)=\prod_j C_j$, which generates a $\Z_2$ group.

To establish the group relation between these two loops, we consider the combined loop
\begin{equation}
    \mathcal V_{\gamma_3}(\theta)=\begin{cases}
\mathcal V_{\gamma_2} (4\theta), & 0\le\theta<\pi/2,\\
\mathcal{V}_{\gamma_1}(4\theta-2\pi)\mathcal V_{\gamma_2} (2\pi), & \pi/2\le \theta<\pi,\\
\mathcal V^{\dagger}_{\gamma_2} (2\theta-2\pi)\mathcal{V}_{\gamma_1}(2\pi)\mathcal V_{\gamma_2} (2\pi), &\pi\le\theta<2\pi.
\end{cases}
\end{equation} 
Thus, $\gamma_3$ consists of the $\gamma_2$ loop, followed by the $\gamma_1$ loop, and finally the inverse of the $\gamma_2$ loop. The corresponding truncated loop with $\theta=2\pi$ in the region $l\le j\le k$ acts on the ground state as
\begin{equation}
\mathcal V^{\text{TR}}_{\gamma_3}(2\pi)|_{\text{G.S.}}=Z_l Z^{-1}_{k},
\end{equation}
 which is precisely the inverse of the action of $\mathcal V^{\text{TR}}_{\gamma_1}(2\pi)$ on the ground state. This establishes the $D_{2n}$ classification of the $S^1$-family.

\subsection{Evolution loops of nontrivial $\text{Rep}(D_{2n})$ NISPT}
When $n=2m$, the $\Z^s_2$-SSB phase admits another unbroken group $K=\Z^{r^2}_{m}\rtimes\Z^{rs}_2$ generated by
\begin{equation}
U_{r^2}=\prod_j X^2_j, U_{rs}=\prod_j X_j C_j\tau^x_j.
\end{equation}
The corresponding NISPT phase is the nontrivial NISPT phase.

A representative $\Z^s_2$-SSB Hamiltonian can be realized as
\begin{equation}
H'_{\Z^s_2\text{-SSB}}=-\sum_j Z^m_j\tau^z_jZ^m_{j+1}\tau^z_{j+1}-\sum_j X_jC_j\tau^x_j+X^{2}_j+X^{-1}_jC_j\tau^x_j+X^{-2}_j.
\end{equation}
Since all terms in this Hamiltonian commute with one another, it has two ground states satisfying the constraint
$Z^m_j\tau^z_j=\pm 1$.
We construct the following evolution loop
\begin{equation}
\mathcal V_{\Z^{rs}_2\text{-charge}}(\theta)=\prod_j\exp\left(\frac{\theta}{2\pi}\text{ln}(Z^{-m}_jZ^m_{j+1})\right),
\mathcal V_{\Z^{r^2}_m\text{-charge}}(\theta)=\prod_j\exp\left(\frac{m\theta Z^m_{j+1}}{4\pi}\text{ln}(Z^{-\tau^z_{j+1}}_jZ^{\tau^z_{j+1}}_{j+1})\right),
\end{equation}
where the second  evolution loop only exists when $m$
is even. Here these two evolutions are both $D_{2n}$ symmetric and pump the $\Z^{rs}_2$ charge and a $\Z_2$ charge of $U_{r^2}$ respectively. 
Together with the evolution loop
$\mathcal{V}_{\mathbb{Z}^s_2\text{-DW}}(\theta)$ that pumps a $\mathbb{Z}_2^s$ domain wall, these loops generate the $S^1$-family. For odd $m$, only the $\mathbb{Z}^{rs}_2$ charge-pumping loop is nontrivial, giving $\mathbb{Z}_2\times\mathbb{Z}_2$
as the $S^1$-family classification. For even $m$, the additional $\mathbb{Z}^{r^2}_m$ charge-pumping loop leads to the non-Abelian $D_8$ structure, and the $S^1$-family is classified by $D_8$.

By the same gauging method, we can obtain the dual  evolution loop
\begin{equation}
\mathcal V_{\gamma_4}(\theta)=\prod_j\exp\left(\frac{m\theta}{2\pi}\text{ln}[Z^{-\mu^x_{j+\frac{1}{2}}}_{j}Z_{j+1}]\right), \quad
\mathcal V_{\gamma_5}(\theta)=\prod_j\exp\left(\frac{m\theta Z^m_{j+1}}{4\pi}\text{ln}[Z^{-\mu^x_{j+\frac{1}{2}}}_{j}Z_{j+1}]\right),
\end{equation}
where the second evolution loop only exists for even $m$.
The corresponding dual Hamiltonian is given by
\begin{equation}
H'_{\text{NISPT}}=-\sum_j Z^m_j\mu^x_{j+\frac{1}{2}}Z^m_{j+1}-\sum_j (X_j+X^{-1}_j)C_j\mu^z_{j-\frac{1}{2}}\mu^z_{j+\frac{1}{2}}+X^{2}_j+X^{-2}_j.
\end{equation}
In particular, the first term imposes the constraint $Z^m_j\mu^x_{j+\frac{1}{2}}Z^m_{j+1}=1$ on the ground state.

Based on the result in the last section, the truncated
$\mathcal V_{\gamma_4}$ with $\theta=2\pi$ in the region $l\le j\le k$ is
\begin{equation}
\mathcal V^{\text{TR}}_{\gamma_4}(2\pi)=Z^{m}_l Z^m_k,
\end{equation}
which forms a $\Z_2$ group on each boundary. We also have $\mathcal{V}_{\gamma_2}(2\pi)=\prod_j C_j$ which forms a $\Z_2$ group.

For even $m$, the evolution loop $\mathcal{V}_{\gamma_5}$ commutes with the first term of the above Hamiltonian. Therefore, the ground state throughout this evolution always satisfies the constraint $Z^m_j\mu^x_{j+\frac{1}{2}}Z^m_{j+1}=1$. As a result, the truncated
evolution loop with $\theta=2\pi$ in the region $ j\le k$ acts on the ground state effectively as
\begin{equation}
\mathcal V^{\text{TR}}_{\gamma_5}(2\pi)|_{\text{G.S.}}=Z^{mZ^{m}_k/2}_k=\sum^{2m-1}_{s_k=0}i^{(-1)^{s_k}s_k}|s_k\rangle\langle s_k|.
\end{equation}
Since $ (\mathcal V^{\text{TR}}_{\gamma_5}(2\pi)|_{\text{G.S.}})^2=Z^m_k$, this loop corresponds to a $\Z_4$ element of $S^1$-family.  Following the same analysis in the previous section, one can then show that the $S^1$-family of the dual phase for even $m$, generated by $\mathcal{V}_{\gamma_2}$ and $\mathcal{V}_{\gamma_5}$, indeed forms a $D_8$ group.
\\
~\\
~\\


\section{Evolution loops of $\Z^o_2\times \Z^e_2$-SSB in a $\Z^o_2\times\Z^e_2\times\Z^a_2$ symmetric systems}
In this section, we construct evolution loops of the $\Z^o_2\times \Z^e_2$-SSB phase in $\Z^o_2\times\Z^e_2\times\Z^a_2$
symmetric system with unbroken group $K=\Z^a_2$, based on the result of the above section. Here we introduce $\mu$ and $\tau$ qubits on each site and define the extended  $\Z^o_2\times\Z^e_2\times\Z^a_2$ symmetry as:
\begin{equation}
 U'_a =\prod_j\exp\left(\frac{\pi i}{4} (1-\sigma^z_{j-1}\sigma^z_j)\right)\mu^x_j,  U_o=\prod_j\sigma^x_{2j-1}\tau^x_j, \quad U_e=\prod_j\sigma^x_{2j}\tau^x_j,
\end{equation} 
which preserves the same mixed anomaly structure as Eq.~\eqref{eq: type3 anomaly ope}. This choice of lattice
realization is also made for the convenience of constructing the noncontractible loops.
Under the KW duality, the  $\Z^o_2\times\Z^e_2\times\Z^a_2$ symmetry is mapped to a $D_8$ symmetry generated by
\begin{equation}
 \tilde{U}_r =\prod_j\exp\left(\frac{\pi i}{4} (1-\tilde{\sigma}^x_j)\right)\mu^x_j, \tilde{U}_s=\prod_j \tilde{\sigma}^z_j\tau^x_j.
\end{equation} 
 Moreover, the $\Z^o_2\times \Z^e_2$-SSB Hamiltonian can be realized by
 \begin{equation}
     H_{\Z^o_2\times \Z^e_2\text{-SSB}}=-\sum_j (\tau^z_j\tau^z_{j+1}+\mu^x_j)+\sum_{j}(\sigma^x_{2j}-\sigma^z_{2j-1}\sigma^x_{2j}\sigma^z_{2j+1}),
 \end{equation}
whose ground states satisfy $\sigma^z_{2j-1}=\pm (-1)^{j}$, $\tau^z_j=\pm 1$ and $-\sigma^x_{2j}=\mu^x_j=1$.

This Hamiltonian is mapped to the $\Z^s_2$-SSB Hamiltonian with unbroken group $K=\Z^r_4$:
 \begin{equation}
     \tilde{H}_{\Z^s_2\text{-SSB}}=-\sum_j (\tau^z_j\tau^z_{j+1}+\mu^x_j)+\sum_j(\tilde{\sigma}^z_{2j}\tilde{\sigma}^z_{2j+1}+\tilde{\sigma}^y_{2j}\tilde{\sigma}^y_{2j+1}),
 \end{equation}
whose ground states satisfy $\tau^z_j=\pm 1$ and $\mu^x_j=1$ with $\tilde{\sigma}^x$ forming a singlet between sites $(2j,2j+1)$.


According to the correspondence in Eq.~\eqref{eq:Z4}, the local Hilbert space of two qubits $\sigma$ and $\mu$ can be mapped to that of a single $\Z_4$ qudit. Then $D_8$ symmetry operator takes the form as Eq.~\eqref{eq:lattice D8 sym} with $n=4$. Therefore, the corresponding nontrivial evolution loops of the above $\Z^s_2$-SSB Hamiltonian are given by Eq.~\eqref{eq:D2n evolution loop}. They can also be represented by
\begin{equation}
\tilde{\mathcal{V}}_{\Z^r_4\text{-charge}}(\theta)=\prod_j\exp\left(\frac{\theta\tau^z_{j+1}}{2\pi}\text{ln}[\tilde{\sigma}^z_j \text{CNOT}_{j}\tilde{\sigma}^z_{j+1}\mu^z_{j+1} \text{CNOT}_{j+1}]\right),\quad \mathcal{V}_{\Z^s_2\text{-DW}}(\theta)=\prod_j\exp\left(\frac{i\theta}{4}(1-\tau^x_j)\right).
\end{equation}
After the inverse KW duality, the first evolution is mapped to
\begin{equation}\label{eq:loop bdw}
    \mathcal{V}_{\gamma_6}(\theta)=\prod_j\exp\left(\frac{\theta\tau^z_{j+1}}{2\pi}\text{ln}[\sigma^x_j \text{CNOT}'_{j}\mu^z_{j+1} \text{CNOT}'_{j+1}]\right)=\prod_j\exp\left(\frac{\theta\tau^z_{j+1}}{2\pi}\text{ln}[ \text{CNOT}'_{j}\sigma^x_j \mu^z_j\mu^z_{j+1}\text{CNOT}'_{j+1}]\right),
\end{equation}
 while the second is invariant.
 
For SSB ground states satisfying $\tau^z_{j}=\pm 1$, we have
\begin{equation}
\begin{split}
 &\mathcal{V}_{\gamma_6}(2\pi)|_{\text{G.S.}}=\prod_j \sigma^x_j,\quad \mathcal{V}_{\Z^s_2\text{-DW}}(2\pi)=\prod_j\tau^x_j.
\end{split}
\end{equation}
Hence, these two evolutions together 
pump the domain walls of $\Z^o_2\times\Z^e_2$ symmetry for SSB ground states. Moreover, when $\theta=4\pi$, truncating the first evolution to the region $ l\le j\le k$ gives
\begin{equation}\label{eq:truncated loop bdw}
    \mathcal{V}^{\text{TR}}_{\gamma_6}(4\pi)=\prod_{l\le j\le k-1}[\sigma^x_j \text{CNOT}'_{j}\mu^z_{j+1} \text{CNOT}'_{j+1}]^{2{\tau^z_{j+1}}}=\prod_{l\le j\le k-1}(\mu^z_{j}\mu^z_{j+1})^{{\tau^z_{j+1}}}.
\end{equation}
For each SSB ground state, this truncated evolution effectively reduces to the boundary operator $\mu^z_l$ and $\mu^{z}_{k}$, each of which carries a $\Z^a_2$  charge. This implies that the first loop is a $\Z_4$ element of the $S^1$-family. Finally, since $\mathcal{V}_{\Z^s_2\text{-DW}}(2\pi)$ maps $ \mathcal{V}_{\gamma_6}(\theta)$ to $ \mathcal{V}_{\gamma_6}(-\theta)$, one can further obtain that the classification is $D_8$ group by the same method as the previous section.

\end{document}